\documentclass{iau_2022}

\usepackage{amsmath}
\usepackage{multirow}
\usepackage{graphicx}
\usepackage{url}
\usepackage{upgreek}

\newcommand{\araa}{Ann. Rev. Astron. Astrophys.}
\newcommand{\apj}{ApJ}
\newcommand{\mnras}{MNRAS}
\newcommand{\aap}{A\&A}
\newcommand{\apjl}{ApJL}

\newcommand{\nat}{Nature}

\begin{document}

\lefttitle{D. Vohl, H.K. Vedantham, J.W.T. Hessels \& C.G. Bassa}
\righttitle{Searching for PRS in dwarf galaxies using LOFAR}

\jnlPage{1}{7}
\jnlDoiYr{2022}
\doival{10.1017/xxxxx}

\aopheadtitle{The Dawn of Cosmology \& Multi-Messenger Studies with Fast Radio Bursts}
\editors{A.C. Editor, E. Keane, K. Rajwade, A. Fialkov \& C. Walker, eds.}
\volno{369}

\title{Searching for FRB persistent radio source counterparts in dwarf galaxies using LOFAR}

\author{
D. Vohl$^{1,2,^*}$, H.K. Vedantham$^{2,3}$, 
J.W.T. Hessels$^{1,2}$ and C.G. Bassa$^2$}

\affiliation{
$^1$\,Anton Pannekoek Institute for Astronomy, University of Amsterdam, \\ P.O. Box 94249, 1090 GE Amsterdam, The Netherlands}

\affiliation{
$^2$\,ASTRON, The Netherlands Institute for Radio Astronomy, \\ Oude Hoogeveensedijk 4, 7991 PD Dwingeloo, The Netherlands
} 

\affiliation{
$^3$\,Kapteyn Astronomical Institute, University of Groningen, \\ P.O. Box 800, 9700 AV Groningen, The Netherlands
}

\affiliation{
$^*$email: {\tt d.vohl@uva.nl}
}

\begin{abstract}
The repeating FRB\,20121102A was localized to a star-forming region in a dwarf galaxy and found to be co-located with a persistent radio source (PRS).  
FRB\,20190520B is only the second known source sharing phenomenology akin to FRB\,20121102A’s, with similar burst activity, host galaxy properties, as well as being associated with a PRS. 
PRS emission is potentially a calorimeter, allowing us to estimate the energy output of the central FRB engine. 
Independently of FRB studies, PRSs have been found in dwarf galaxies and interpreted as intermediate mass black holes.
To improve our understanding of such sources associated with dwarf galaxies, it is imperative to increase the known sample size.
Here, we present a search for compact radio sources coincident with dwarf galaxies, discuss source candidates and planned strategies to differentiate them between candidate FRB hosts and intermediate mass black holes.
\end{abstract}


\maketitle

\section{Introduction}
\label{sec:introduciton}

FRB\,20121102A was the first fast radio burst (FRB) source found to repeat~\citep{2016Natur.531..202S}, ruling out progenitor models related to cataclysmic scenarios at least for some FRBs.
Repeat bursts provided the means for the first precise localisation of an FRB to a host galaxy~\citep{2017Natur.541...58C}. 
The galaxy hosting this source is a low-mass, low-metallicity dwarf galaxy at redshift $z=0.193$, which is a typical host for long gamma-ray bursts and super-luminous supernovae~\citep{2017ApJ...834L...7T}.
These facts hinted that the FRB engine may be born in such explosions.
Moreover, FRB\,20121102A was the first FRB to be co-localised with a persistent radio source~\citep[PRS;][]{2017Natur.541...58C}, with luminosity $L_\mathrm{radio}\sim10^{39}\,\mathrm{erg\,s^{-1}}$.

Very long baseline interferometry (VLBI) observations showed that the FRB and the PRS are located within $40\,{\rm pc}$ of one another~\citep{2017ApJ...834L...8M}, unambiguously connecting the two.
The PRS has a flat spectral index
$S_\nu\propto\nu^\alpha$, with $\alpha=-0.27\pm0.24$. Bursts from FRB\,121102A have high and variable rotation measure~\citep{2018Natur.553..182M}, and the dispersion measure (DM) has been shown to increase over time~\citep{2019ApJ...876L..23H, 2021MNRAS.505.3041P}. 
Finally, the spectral energy distribution---derived from multi-wavelength measurements and upper limits---matches that of the Crab nebula, though with orders of magnitude higher luminosity.
{ The combination of these measurements make a plausible case for an FRB engine that is a young, highly magnetized neutron star that is embedded in an expanding supernova remnant and powering a wind nebula (plerion).} 
Another plausible explanation is that it is a low-luminosity accreting massive black hole. 
{ Incidentally, \citet{2020ApJ...888...36R} have identified a sample of PRSs associated with dwarf galaxies and suggested that they may be the long-sought population of intermediate mass black holes (IMBHs) that were predicted to reside in dwarf galaxies. 
Although it is unclear if IMBHs are related to FRBs---most FRB models prefer a magnetar progenitor~\citep[e.g., see discussion in][]{2020ApJ...895...98E}---PRSs in dwarf galaxies are an interesting radio source population in their own right.} 

Only recently was another repeating FRB co-localised to a PRS reported~\citep[][]{2022Natur.606..873N}. Like FRB\,20121102A, FRB\,20190520B is hosted in a star-forming dwarf galaxy, has a large $\mathrm{DM}_\mathrm{host}$ contribution, has a high repetition rate, and the PRS has a shallow spectral index ($\alpha=-0.41\pm0.04$).
{ Given these two cases of FRB/PRS connection, it seems that PRSs represent an important aspect of some FRBs, although to what extent remains an open question.
For instance, if PRSs are wind nebulae, a limited lifespan during which they could be detected~\citep[$\sim$few centuries;][]{2006ARA&A..44...17G} could explain why only a subset of FRBs have a PRS counterpart.
Considering a sample of 15 localised FRBs having radio sensitivity limits that could allow to detect a PRS, where six out of 15 are repeaters, \citet{2022ApJ...927...55L} estimate that PRS occurrence could be as high as 20\% for repeating FRBs given that 2 out of the 6 repeating FRBs are associated to a PRS. Furthermore, these authors estimate that PRS measurements should not be strongly biased by distance.}

To improve our understanding of PRSs and their potential connection to the FRB and/or IMBH phenomena, it is imperative to increase the known sample size.
Here, we present a targeted search for compact radio sources coincident with dwarf galaxies, using the LOFAR Two-Meter Sky Survey (LoTSS) second data release~\citep[DR2;][]{2022A&A...659A...1S}---the most sensitive large-area survey for optically thin synchrotron emission---as our radio reference catalog. 

\section{Searching for PRS candidates}
\label{sec:search}

LoTSS DR2~\citep{2022A&A...659A...1S} comprises about 4 million radio sources spanning over $5600~{\rm deg}^2$ of the northern sky. 
LoTSS operates at a central frequency of 144\,MHz with 48 MHz of bandwidth (120--168MHz).
The survey has a  $\sim6\times({\small \frac{144\,\rm{MHz}}{v}})~\rm{arcsec}$ resolution and a median root mean square sensitivity of about $80\,\rm{\upmu Jy/beam}$.
Furthermore, with a 0.2 arcsec astrometric uncertainty\footnote{This is rms deviation of LoTSS sources from their PanSTARRS counterparts. The aggregate astrometric error for any given LoTSS sources also depends on its SNR; see \S 3.2 of \citet{2022A&A...659A...1S} for details.} for sources brighter than 20~mJy, comparable to optical surveys, and a 90\% point source completeness for sources $\geq0.8$ mJy/beam, LoTSS DR2 represents an excellent catalog for our study.

We cross-match LoTSS DR2 to the Census of the Local Universe \citep[CLU;][]{2019ApJ...880....7C} catalogue.
CLU is a photometric survey carried out in four H$\upalpha$ bands corresponding to redshift{ s} up to $z=0.0471$. 
CLU comprises about 270,000 sources spanning 3$\pi$ of the sky, of which over 82,000 fall within the LoTSS DR2 footprint.
The CLU catalog includes various derived physical parameters such as the stellar mass, star formation rate (SFR), and redshift.  
We constrain these sources to a mass range corresponding to dwarf galaxies ($10^7 \leq M_\star/M_\odot \leq 3\times10^9$ ), leaving 26,324 sources that form our parent sample. 

We apply a moderate peak brightness cut at 0.8 mJy/beam to LoTSS DR2 sources to limit our sample to the 90\% complete point-source sample---leaving 2,622,903 sources. 
We cross-match object coordinates from each catalog, taking into account astrometric uncertainties provided in DR2, using a 6 arcsec radius (the resolution of LoTSS), yielding 2161 matches. 
We constrain the matched sample to reject extended sources using the $R_{99.9}$ compactness criterion~\citep[see][Eqn. 2]{2022A&A...659A...1S}, where the ratio of the natural logarithm of the integrated flux density ($S_I$) to peak brightness ($S_P$) is less or equal than the envelope that encompasses the 99.9 percentile of the $S_I/S_P$ distribution---leaving 869 sources. 

\begin{figure}
\includegraphics[width=13.4cm]{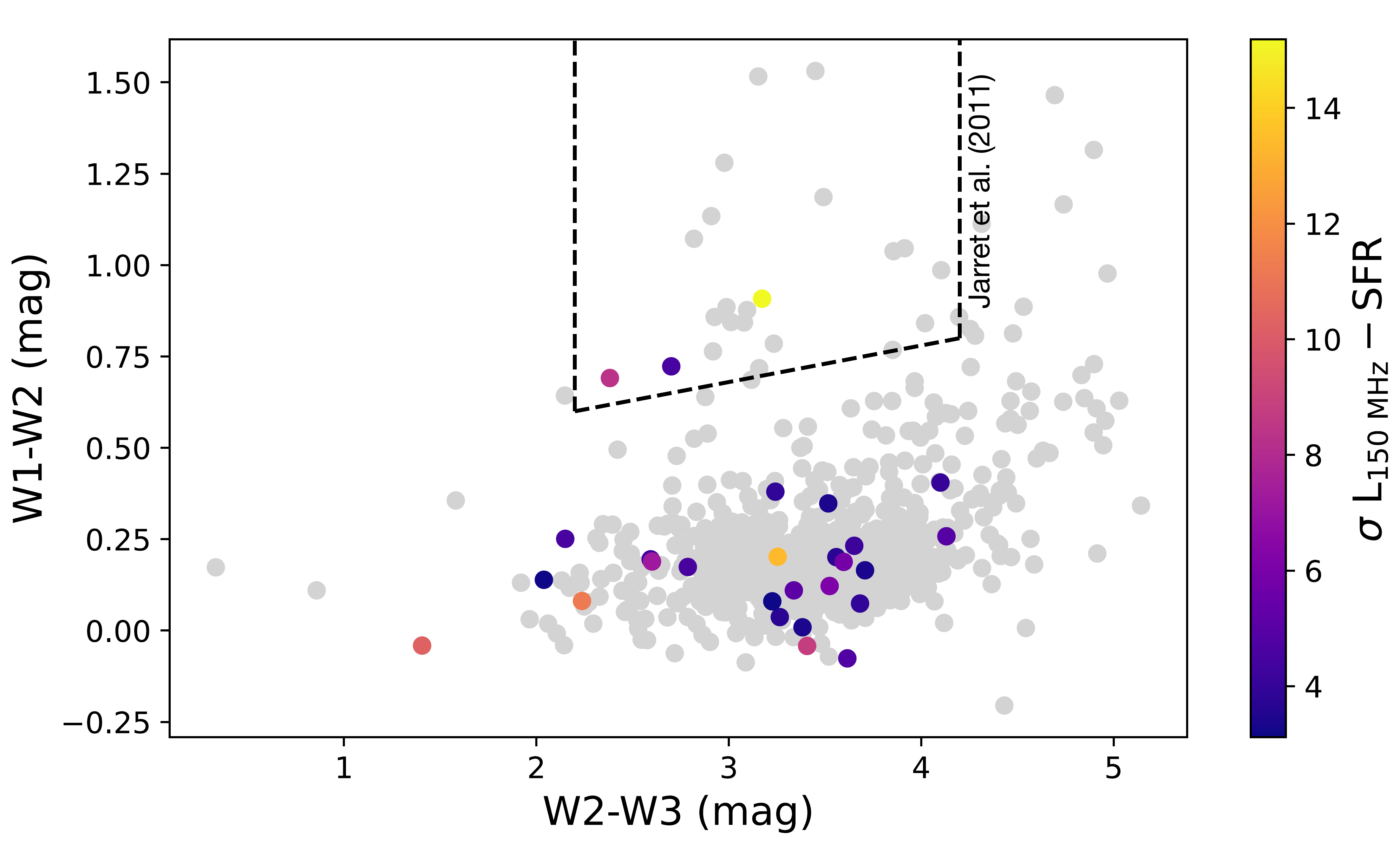}
\caption{{ WISE color-color plot.} Coloured markers correspond to dwarf galaxies matched to a compact radio source with luminosity exceeding $3\sigma$ in the L--SFR relation, with the colour indicating its $\sigma$ value via the colour bar. White and grey-filled markers indicate all galaxies and dwarf galaxies matched to a radio source, respectively. We remove from our sample the three sources falling within the AGN region marked by dashed lines.}
\label{fig:wise}
\end{figure}

We then constrain these compact sources to those exceeding $3\sigma$ relative to the luminosity-SFR (L--SFR) relation presented by \citet{2018MNRAS.475.3010G}, which yields 27 sources. 
We note that out of the subset of 869 compact sources, 263 do not have SFR information in CLU. 
Next, we remove obvious active galactic nuclei (AGNe) from the set. 
We do this filtering by employing WISE colors (Figure \ref{fig:wise}). Using WISE filters W1, W2, and W3, we exclude sources falling within the AGN region prescribed by \citet[][Eqn. 1]{2011ApJ...735..112J} in the W1-W2 and W2-W3 plane.
Three candidates fall within this region, leaving 24 sources. 

\begin{figure}
\includegraphics[width=13.4cm]{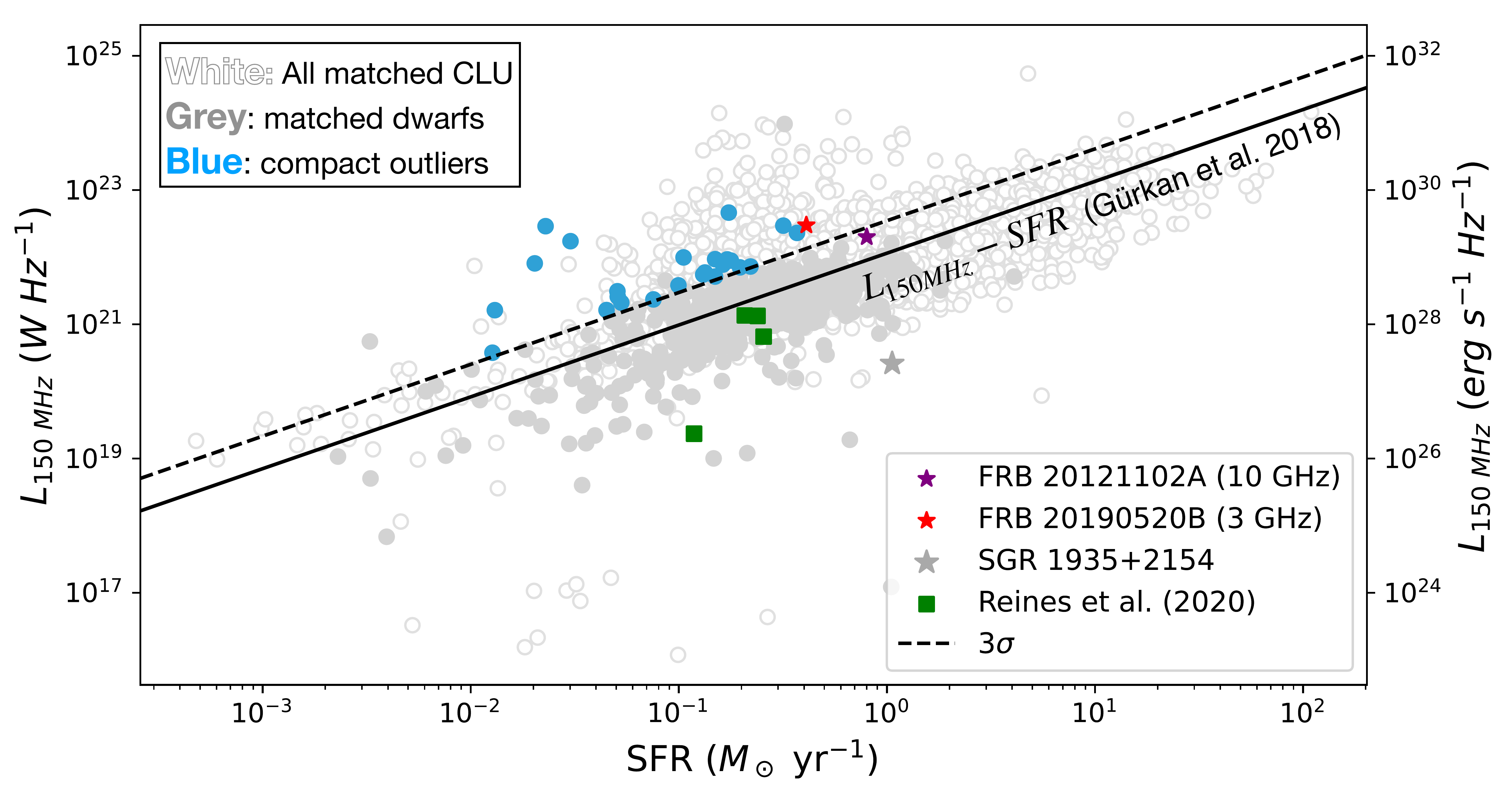}
\caption{{ Candidate selection via the luminosity--star formation rate relation.} White and grey-filled markers indicate all galaxies and dwarf galaxies matched to a radio source, respectively. Blue circles correspond to our final selection of compact radio sources matched to a dwarf galaxy with luminosity exceeding $3\sigma$ (dashed line) on the L-SFR relation by \citet[][solid line]{2018MNRAS.475.3010G}. As reference, we show the PRS luminosity for both FRB~20121102A~\citep{2017ApJ...834L...7T, 2022ApJ...927...55L} and FRB~20190520B~\citep{2022Natur.606..873N}, assuming flat spectral indices. 
Finally, we show \citet{2020ApJ...888...36R} galaxies matched in CLU for which SFR information is available, and SGR~1935+2154 using published FRB-like event's luminosity and SFR values from \citet{2021MNRAS.501.3155W} and \citet{2010ApJ...710L..11R}, respectively.}
\label{fig:selection}
\end{figure}

Finally, we verify that redshifts listed in CLU match that of associated sources listed in NED as redshifted emission lines other than H$\upalpha$ could fall within CLU’s filter and masquerade as H$\upalpha$. 
We note that all redshifts for this subset are correct. 
We depict the candidate selection on the L--SFR plane in Figure \ref{fig:selection}.

\section{Candidates}
\label{sec:candidates}

\begin{figure}
\centering
\begin{tabular}{cc}
\includegraphics[width=45.5mm]{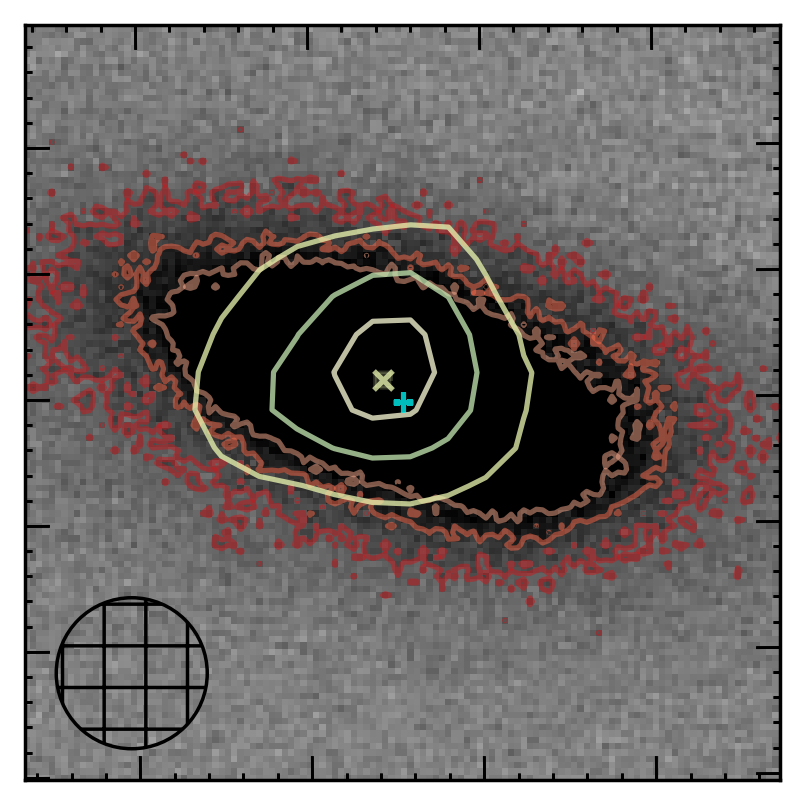} &  
\includegraphics[width=45.5mm]{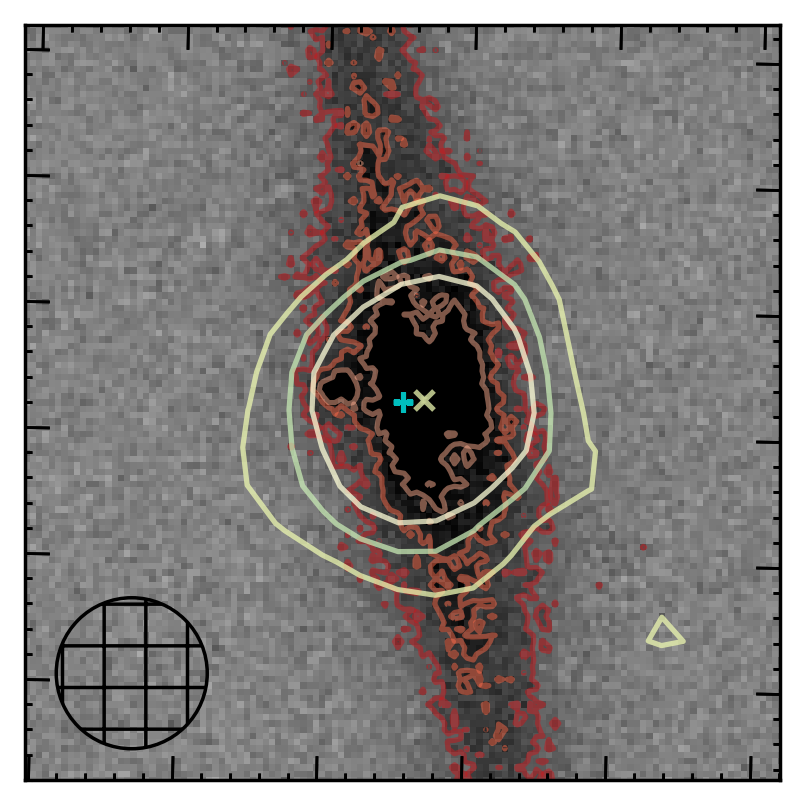} \\
\includegraphics[width=45.5mm]{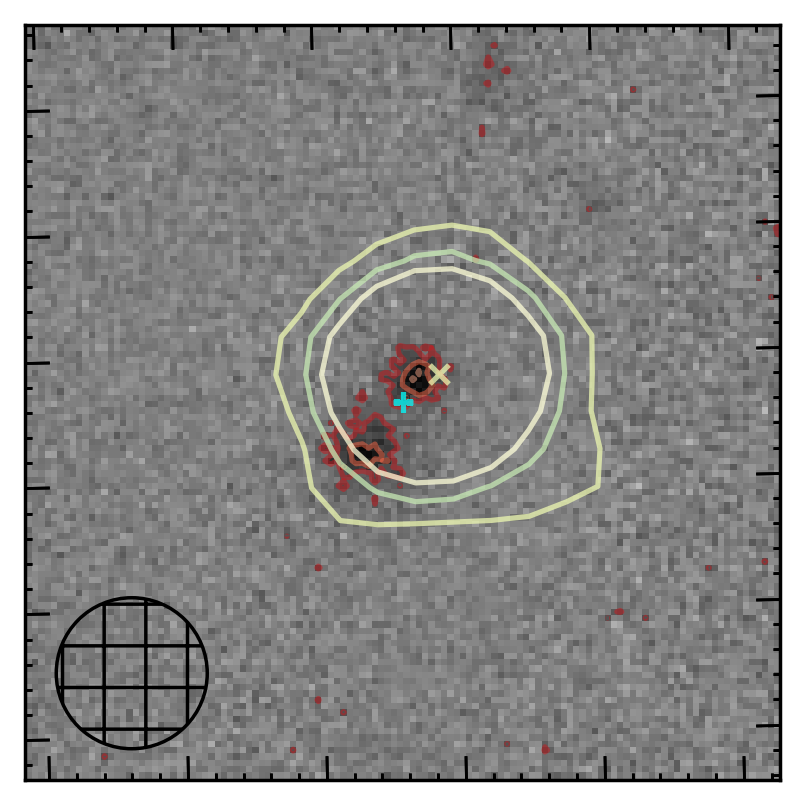} &
\includegraphics[width=45.5mm]{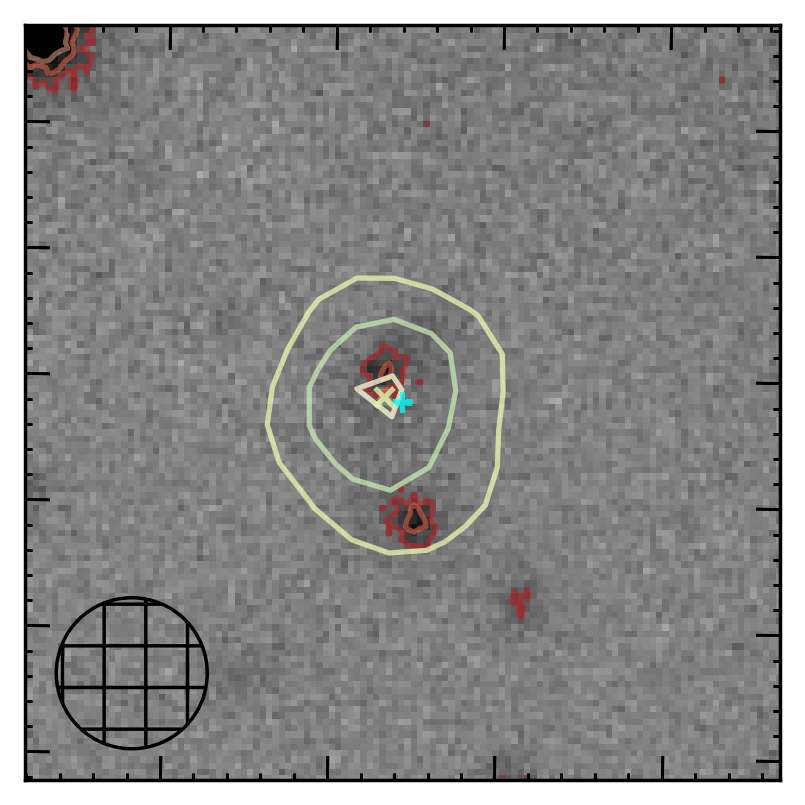} \\
\includegraphics[width=45.5mm]{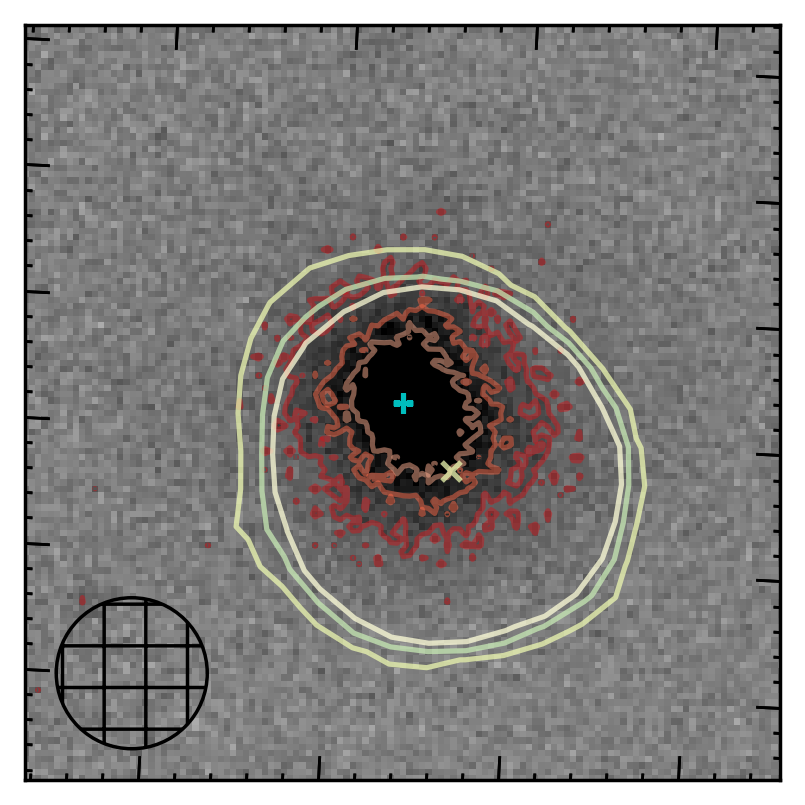} &
\includegraphics[width=45.5mm]{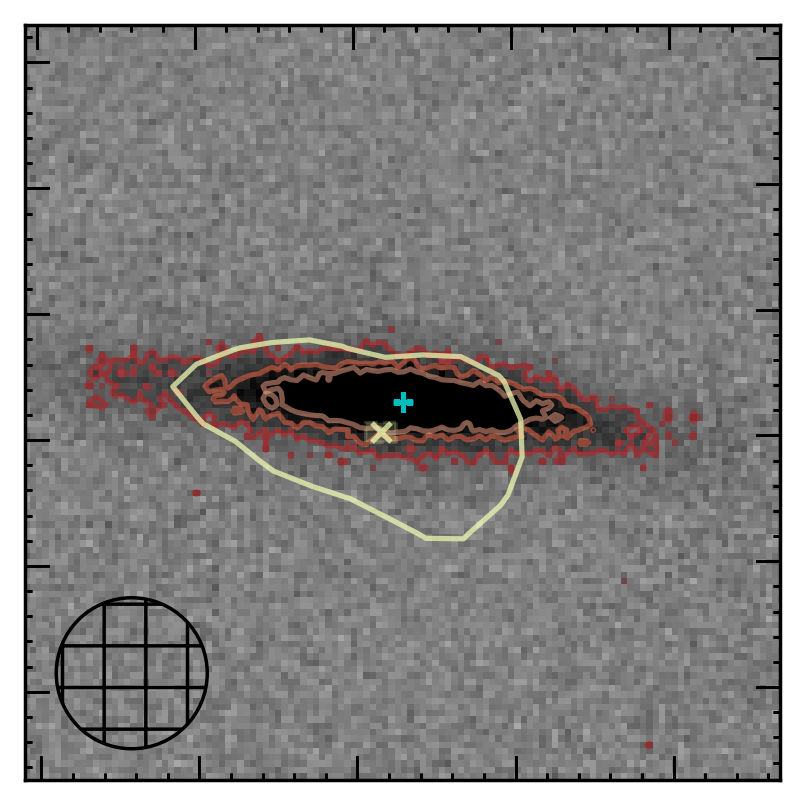} \\
\includegraphics[width=45.5mm]{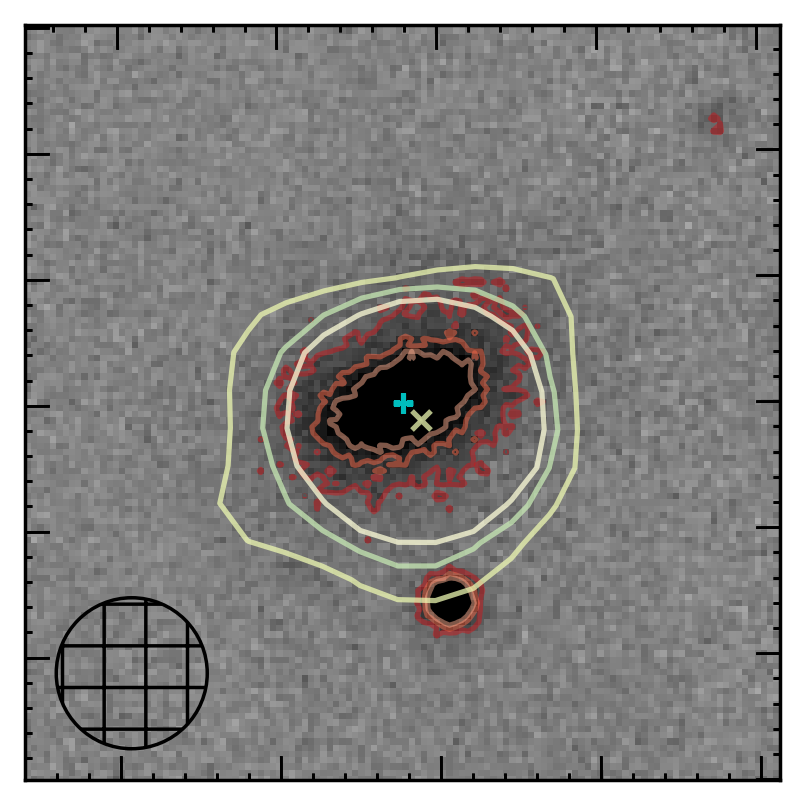} &
\includegraphics[width=45.5mm]{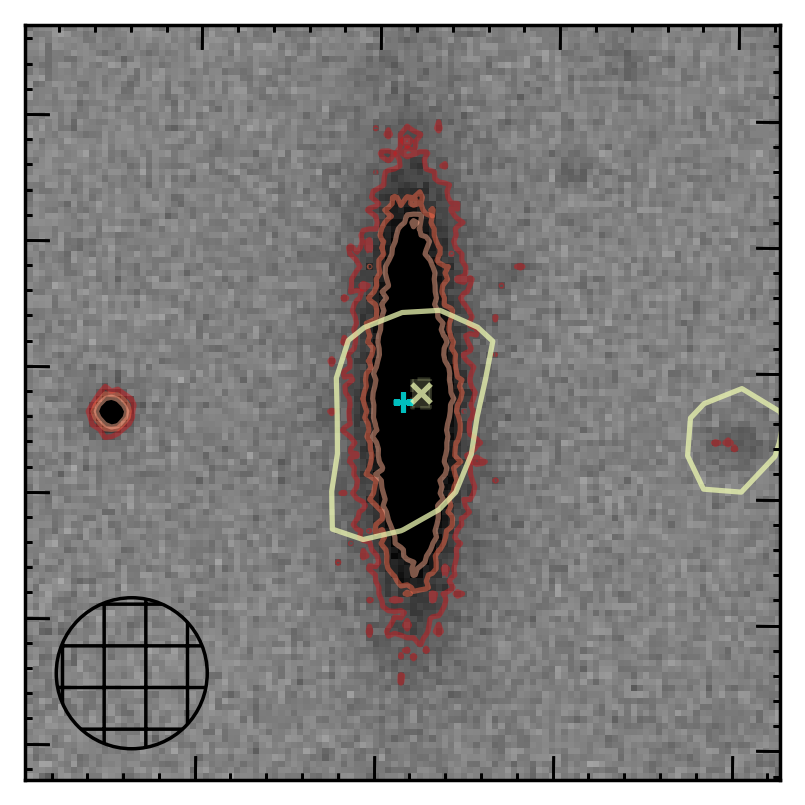} \\
\end{tabular}
\caption{{ PRS candidates.} Each panel shows a PanSTARSS R filter image in grey scale set to the median $\pm~15$ median absolute deviation. 
Contours show optical data from PanSTARRS (red) and radio data from LoTSS (yellow) at 5, 10, and 15$\sigma$ levels. 
The yellow (x) and blue (+) markers show the central coordinates of matched objects from LoTSS and CLU, respectively. 
LoTSS astrometric uncertainty is marked as a yellow box. 
Black hashed circles indicate the 6'' beam shape from LoTSS.}
\label{fig:images}
\end{figure}

Figure \ref{fig:images} shows composite images for a subset of our source candidates, where we show radio contours from LoTSS DR2 in yellow, optical contours in red from the r filter from PanSTARRS~\citep{2016arXiv161205560C}, which includes all four H$\upalpha$ bands from CLU.
Contours are set to 5$\sigma$, 10$\sigma$ and 15$\sigma$, respectively. 

Given the CLU optical catalog specificity, all 24 candidates are within the local Universe ($z\leq0.04$), with distances ranging from about 22 to 160\,Mpc. 
Their stellar masses range from $10^{7.9}$ to $10^{9.4}\,M_\odot$, SFR between 0.01 to 0.37 $M_\odot\,\rm{yr}^{-1}$, and specific SFR (SFR per unit stellar mass) range from 0.01 to 0.49 $\rm{Gyr}^{-1}$.
The projected offset between the optical and radio coordinates for these candidates range between about 0.19 to 7 arcsec, taking into account positional uncertainties for the radio sources---explaining why some sources go beyond the 6 arcsec cut used during cross-matching. 
Given the matched galaxies' redshifts, these offsets correspond to a range of about 37 to 4669 parsec. 
Out of the 24 source candidates selected through the method described in Section \ref{sec:search}, nine sources are matched in other radio surveys, including nine matched at 1.4\,GHz in the VLA FIRST survey~\citep{1995ApJ...450..559B}, and four at 3\,GHz in the VLASS Epoch 1 Quick Look Catalogue~\citep{2020RNAAS...4..175G}, which appear as unresolved point sources in images from both these surveys. 
In addition to the four matches from VLASS Epoch 1 Quick Look Catalogue, six extra sources appear as point source in images that can be queried via \url{http://cutouts.cirada.ca} that are not yet part of the public catalogue. 
Initial results fitting a single power-law model for these sources from 144\,MHz to 3\,GHz (where applicable) yields spectral indices varying between $-1.01$ to 0.28. 

\section{Closing remarks}
\label{sec:closing}

We are currently investigating this set of candidates to better characterize their nature. 
In particular, we are gathering spectra to evaluate whether star formation or AGN activity is the dominant source of ionization within these galaxies. 
Furthermore, a critical step towards establishing them as potential FRB hosts is to conclusively determine the compactness of these sources. 
For this purpose, we have requested observing time with the European VLBI Network, as well as computing time to re-image the archival LoTSS radio data on these sources with inclusion of LOFAR's international stations that will give us a resolution of about $0''.25$ (LOFAR-VLBI).
{ We also plan to search these targets for millisecond-duration bursts. 
Starting with the hypothesis that some are similar in nature to currently known PRSs, we can expect these to be repeating FRB sources.
Furthermore, given the periodic activity of some FRBs like FRB\,20121102A~\citep{2020MNRAS.495.3551R} and FRB\,20180916B~\citep{2020Natur.582..351C}, it is plausible that a subset of our candidates can also display on/off phases of FRB emission.}
We end by noting that due to LoTSS' unprecedented sensitivity to optically thin synchrotron sources in a wide-angle survey, we have been able to select interesting radio sources in dwarf galaxies. 
The proposed VLBI observations are a crucial step towards the discovery of a new population of either wind nebulae or black-holes in nearby dwarf galaxies---both outcomes being scientifically interesting.

\end{document}